\documentclass[sigconf]{acmart}

\copyrightyear{2026}
\acmYear{2026}
\setcopyright{cc}
\setcctype{by}
\acmConference[UIST '26]{The 39th Annual ACM Symposium on User Interface Software and Technology}{November 02--05, 2026}{Detroit, MI, USA}
\acmBooktitle{The 39th Annual ACM Symposium on User Interface Software and Technology (UIST '26), November 02--05, 2026, Detroit, MI, USA}
\acmDOI{10.1145/3830398.3830581}
\acmISBN{979-8-4007-2856-3/2026/11}

\begin{document}
\raggedbottom

\title{Electrotactile Improves Thermal Referral}

\author{Wen Li}
\orcid{0009-0007-4512-0240}
\affiliation{%
  \institution{The University of Chicago}
  \city{Chicago}
  \state{Illinois}
  \country{USA}}
\email{wenli@uchicago.edu}

\author{Rong Ni}
\orcid{0009-0008-2869-0580}
\affiliation{%
  \institution{The University of Chicago}
  \city{Chicago}
  \state{Illinois}
  \country{USA}}
\email{rongn@uchicago.edu}

\author{Bozhi Tian}
\orcid{0000-0003-0593-0023}
\affiliation{%
  \institution{The University of Chicago}
  \city{Chicago}
  \state{Illinois}
  \country{USA}}
\email{btian@uchicago.edu}

\author{Pedro Lopes}
\orcid{0000-0001-6527-7084}
\affiliation{%
  \institution{The University of Chicago}
  \city{Chicago}
  \state{Illinois}
  \country{USA}}
\email{pedrolopes@uchicago.edu}

\renewcommand{\shortauthors}{Li et al.}

\begin{abstract}
Thermal referral enables thermal sensations in locations lacking thermal actuators---this is achieved using vibrotactile actuators to redirect a nearby thermal sensation to where a tactile sensation is applied. However, we found that its reliance on vibration introduces critical limitations: it struggles to produce cold referral, and the inherent strong tactile ``buzz'' makes it unsuitable for simulating non-contact thermal events, such as the chill of an open freezer in VR (in contrast to contact-based thermal events like touching the freezer's cold handle). To improve this, we propose a shift from vibrotactile to electrotactile-based thermal referral. We evaluated in two user studies---a psychophysics experiment (N=22) and a VR deployment (N=20)---where we contrasted electrotactile with vibrotactile-based thermal referral. Our results reveal key advantages of the electrotactile-based thermal referral: (1) increases the referral rate for cold sensations; (2) increases thermal perception while minimizing tactile; and (3) improves realism across a range of VR thermal scenarios, specifically distinguishing between contact-based and non-contact thermal events. Finally, we provide design guidelines for choosing tactile cues to create immersive multi-modal thermal experiences in VR.
\end{abstract}


\begin{teaserfigure}
  \includegraphics[width=\textwidth]{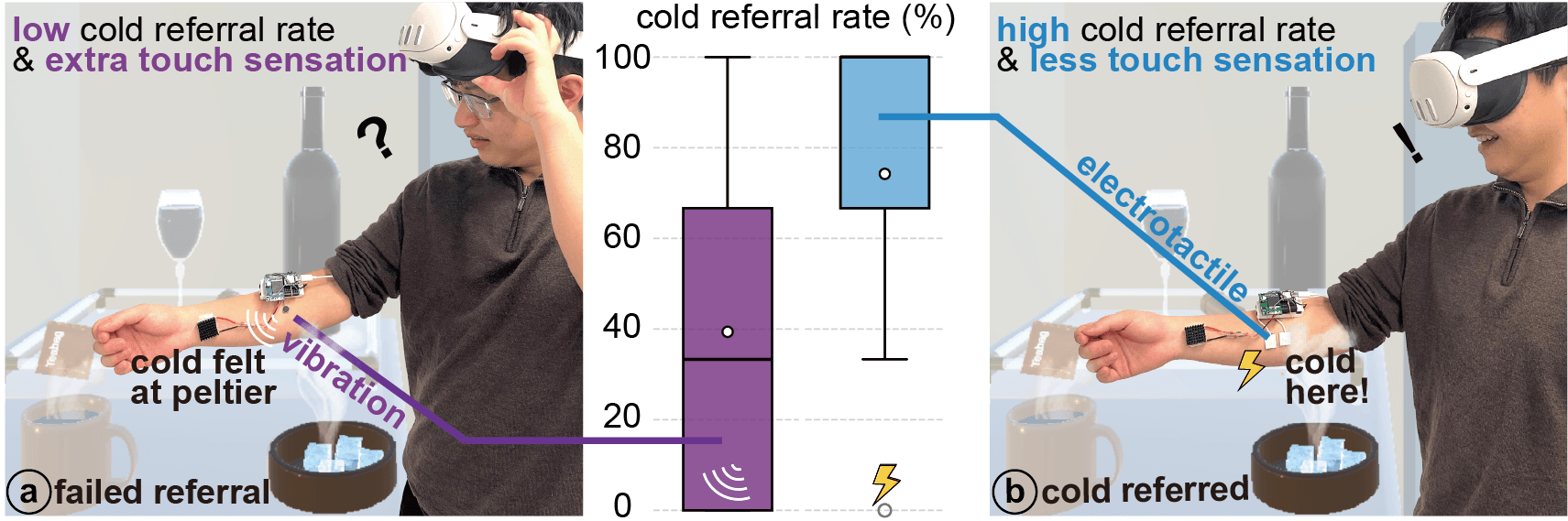}
  \Description{Composite figure showing a vibrotactile thermal-referral failure and an electrotactile success. On the left, a person in VR feels cold only at the wrist Peltier rather than near the elbow tactile site. The center box plot shows a higher cold-referral rate for electrotactile than vibrotactile. On the right, electrotactile refers cold to the intended location near the elbow.}
  \caption{Thermal referral emerged as a popular technique to minimize the number of thermal actuators needed by swapping some with smaller and less power-hungry vibrotactile actuators. In this technique the vibration motors redirect nearby thermal sensations to the vibration motor's location. However, despite the rapid expansion of this technique in HCI, we unveiled critical challenges that limit its interactive applicability: (1) thermal referral struggles to reliably produce cold referral, and (2) the distracting tactile ``buzz'' from vibration motors makes it unsuitable for simulating non-contact thermal sources. Thus, we propose a methodological shift from vibrotactile to \emph{electrotactile} stimulation. Our results confirmed this significantly improves cold referral rates and thermal quality.}
  \label{fig:teaser}
\end{teaserfigure}

\maketitle

\hypertarget{introduction}{%
\section{INTRODUCTION}\label{introduction}}

Spatially distributed thermal feedback can make wearable and immersive interfaces more expressive, but scaling such feedback typically requires adding thermal actuators at every possible output location. Because these actuators are comparatively bulky and power-hungry, increasing their number limits wearability. \textit{Thermal referral} offers an alternative: when tactile stimulation is applied near a thermal actuator, users can perceive the temperature at the tactile location rather than at its physical source~\citep{Wang2024Thermal}. By decoupling the physical and perceived locations of thermal output, this illusion enables interfaces to cover more body locations with fewer thermal actuators.


Recent interactive systems have begun using thermal referral as an actuation technique for localized, moving, and object-coupled thermal feedback in wearable and immersive interfaces \citep{Cataldo2016Thermal,Singhal2025HeatFlow,Singhal2024Thermal,Son2023Upper,Wang2024Thermal,Wang2024Fiery,Yi2024ThermicVib}. Across these systems, the standard implementation is based on vibrotactile actuators (e.g., ERM or LRA motors) to anchor the referred thermal sensation. 

While effective
for specific scenarios---such as simulating contact heat where a user
expects a physical contact sensation (e.g. feeling hot water splash
one's hand~\citep{Wang2024Fiery})---we will demonstrate, via the findings of our two
experiments, that relying on mechanical vibration introduces two critical
limitations that hinder broader application of thermal referral in
interactive domains, such as VR. First, vibrotactile thermal referral is unreliable for cold output: prior work found its occurrence rate to be approximately 20\% lower for cold than for heat \citep{Wang2024Thermal}. In an interactive system, such failures produce a mismatch between the intended output location and the user’s perception, breaking immersion. Second, vibration adds an unavoidable tactile “buzz” to the thermal percept. Although this cue can be congruent with contact events, such as touching a hot object, it conflicts with non-contact events, such as feeling heat from a fireplace or cold air from an ice source, as shown in Figure~\ref{fig:teaser}. Thus, vibrotactile referral cannot cleanly distinguish thermal events that imply contact from those that do not.

We overcome these limitations by proposing a novel thermal referral via
\emph{electrotactile stimulation,} as depicted in Figure~\ref{fig:teaser} (b). We evaluated this approach in two studies. Our first study providerd a controlled psychophysics setting (\emph{N}=22) that compared vibrotactile and electrotactile referral across modality and intensity, demonstrating that electrotactile stimulation substantially improves cold referral and that threshold-level stimulation can reduce the accompanying tactile sensation (our vibrotactile baseline also provides a partial replication of prior thermal-referral findings~\citep{Wang2024Thermal}). Our second study (\emph{N}=20) examined how these perceptual differences affect four contact and non-contact thermal interactions in VR. Together, this work contributes: (1) electrotactile thermal referral as an alternative actuation technique for spatial thermal interfaces; (2) a wearable, interactive implementation and empirical characterization of tactile modality and intensity; and (3) design guidance for matching tactile cues to the semantics of contact and non-contact thermal events.

\hypertarget{related-work}{%
\section{RELATED WORK}\label{related-work}}

The work presented in this paper builds primarily on the broadly defined
field of haptics (i.e., thermal included), with particular emphasis on
the emergence of thermal referral in HCI as a promising technique for
thermal rendering. Finally, since our approach substitutes mechanical
vibration with electrotactile, we succinctly review the field of
electrotactile.

\hypertarget{thermal-referral-effect}{%
\subsection{Thermal referral effect}\label{thermal-referral-effect}}

A \emph{thermal referral} is a type of illusion that perceptually
displaces a thermal sensation to the location where a tactile stimulus
is delivered---i.e., the thermal sensation is thus \emph{referred} to
the location of the tactile stimulus, giving rise to the term ``thermal
referral''~\citep{Green1977Localization,Ho2011Mechanisms}. This effect recently rose to prominence in
HCI~\citep{Cataldo2016Thermal,Singhal2025HeatFlow,Singhal2024Thermal,Son2023Upper,Wang2024Thermal,Wang2024Fiery,Yi2024ThermicVib} by providing a powerful way to minimize the
number of \emph{thermal} actuators (i.e., by swapping some thermal
actuators with smaller and less power-hungry tactile actuators). This
technique is in line with a larger trend in haptics arguing for less
cumbersome actuators, without sacrificing their actuation
capabilities~\citep{brooksTrigeminalbasedTemperatureIllusions2020,Pacchierotti2017Wearable,Teng2026Next}. Thermal referral has thus been explored to
enable more and energy-efficient thermal interfaces~\citep{Wang2024Fiery}.

\hypertarget{vibration-dominates-thermal-referral}{%
\subsection{Vibration dominates thermal referral}\label{vibration-dominates-thermal-referral}}

While the choice of tactile actuators for thermal referral has varied
slightly, the vast majority uses \emph{mechanical stimulation}, with
vibration motors (e.g. linear resonant actuators or eccentric rotating
mass motors) as the standard actuator for thermal referral
HCI~\citep{Liu2021ThermoCaress,Singhal2025HeatFlow,Singhal2024Thermal,Son2023Upper,Wang2024Thermal,Wang2024Fiery,Yi2024ThermicVib}.

Examples of vibrotactile-based thermal referral include: Son et al.'s
upper body thermal referral vest using vibrotactile masking to induce
localized thermal feedback on the torso~\citep{Son2023Upper}; \emph{Thermal
Masking}~\citep{Wang2024Thermal}, a systematic investigation of vibrotactile-based
thermal referral; \emph{Fiery Hands}~\citep{Wang2024Fiery}, a thermal glove
integrating vibrotactile and thermal feedback for virtual object
manipulation; \emph{Thermicvibro}~\citep{Yi2024ThermicVib}, a multimodal haptic glove
coupling vibration with thermal actuation to enable dynamic thermal
sensations during VR interaction; and \emph{Thermal In Motion}~\citep{Singhal2024Thermal}
and \emph{HeatFlow}~\citep{Singhal2025HeatFlow} leveraging thermal referral to create
moving thermal cues.

While other mechanical actuators such as microblowers
(\emph{ThermoCaress}~\citep{Liu2021ThermoCaress}) have been used in thermal referral, these
are strikingly less common than vibrotactile actuators. This is likely a
consequence of vibrotactile stimulation being a popular choice for
rendering touch sensations, with vibration motors commonly integrated
across haptic devices like vests~\citep{Lindeman2004Towards,Lindeman2006Wearable,Son2023Upper} or gloves~\citep{Wang2024Fiery,Yi2024ThermicVib}.

\hypertarget{limits-of-vibrotactile-based-thermal-referral}{%
\subsection{Limits of vibrotactile-based thermal
referral}\label{limits-of-vibrotactile-based-thermal-referral}}

Implementing thermal referral with vibration introduces psychophysical
limitations. First, the reliability of the illusion is asymmetric. Prior
work noted that while heat is referred reliably, cold referral rates are
strikingly 20\% lower~\citep{Wang2024Thermal}. The prevailing hypothesis attributes
this to the distribution of thermoreceptors. As cold receptors outnumber
warm receptors, the thermal redistribution among thermal and tactile
actuator areas is thought to be less effective~\citep{Ho2011Mechanisms,Luo2020High}. While this
could account for some differences between hot and cold referral, our
findings will demonstrate it might not be a complete explanation.

Second, vibration motors for thermal referral create feelable mechanical
vibrations on the skin, i.e., they ``buzz'' and create a sensation that
users interpret as a ``touch''. While this extra tactile percept is
acceptable for simulating contact thermal events (e.g., touching a warm
cat~\citep{Yi2024ThermicVib} or hot water~\citep{Wang2024Fiery}), it creates a sensory conflict when
simulating non-contact thermal events (e.g., ambient/radiant thermal
sources like an open freezer or a fireplace). Our work addresses these
issues by replacing the vibration actuators with electrotactile
stimulation.

\hypertarget{electrotactile-stimulation}{%
\subsection{Electrotactile
stimulation}\label{electrotactile-stimulation}}

Electrotactile stimulation renders tactile sensations by passing
currents via the skin. It has been widely explored for its ability to
deliver tactile sensations in extremely compact form factors, such as
fingertip displays~\citep{Kajimoto2012Electrotactile,Lin2022Super,Teng2024Haptic,Withana2018Tacttoo}, sensory
substitution~\citep{Franceschi2017System,Teng2025Seeing,Uematsu2016Tactile}, and popularly, to render touch in virtual
environments~\citep{Huang2023skin,Shi2021Self,Tanaka2023Full,Ushiyama2023FeetThrough,Vizcay2023Design,Yem2017Wearable}. Unlike vibration, electrotactile
stimulation has no moving parts (enabling extremely thin form
factors~\citep{Teng2024Haptic,Withana2018Tacttoo}) and generates almost no heat loss (hailed for its
extremely low-power consumption compared to vibrotactile~\citep{Shi2021Self}).

Importantly, the sensations induced by electrotactile can be modulated
by varying parameters such as pulse width, frequency, intensity, and
polarity \citep{Alotaibi2022First,Djozic2015Psychophysical,Kaczmarek1991Electrotactile,Kajimoto2012Electrotactile,Kourtesis2022Electrotactile}. Variations of these enable selective
stimulation of many mechanoreceptors (including Merkel disks, Meissner
corpuscles, Pacinian corpuscles and Ruffini endings)~\citep{KajimotoElectro}, and a
range of sensations from subtle tingling to pressure~\citep{Djozic2015Psychophysical}. Prior
studies have reported electrotactile-induced cold sensations under
specific conditions. Electrotactile stimulation induced cold sensations
on the forehead \citep{Saito2021Thermal}, potentially because the forehead has greater
cold sensitivity than the forearm \citep{Seo2021Differential}. Another study extended this
effect to the arm by applying gel or anesthetics \citep{Saito2023Coldness}, potentially
by activating myelinated A-$\delta$ afferents and suppressing competing
sensations mediated by other nerve fibers. These studies demonstrated
that electrotactile stimulation can directly induce cold sensations
under specific conditions. But as we note in our study, none of our
participants reported feeling cold sensations during the extensive
electrotactile calibrations. One prior work has combined electrotactile
stimulation with thermal actuators in a single device for delivering
co-located thermal/tactile feedback~\citep{Huang2023skin}. However, electrotactile
has never been investigated as a tactile source for thermal
referral---this is one of our contributions. We hypothesized that the
ability of electrotactile stimulation to recruit a range of
mechanoreceptors offers an alternative pathway to trigger thermal
referral that does not rely on mechanical vibrations.

\hypertarget{implementation-electrotactile-thermal}{%
\section{Implementation: electrotactile +
thermal}\label{implementation-electrotactile-thermal}}

\begin{figure}
  \centering
  \includegraphics[width=\linewidth]{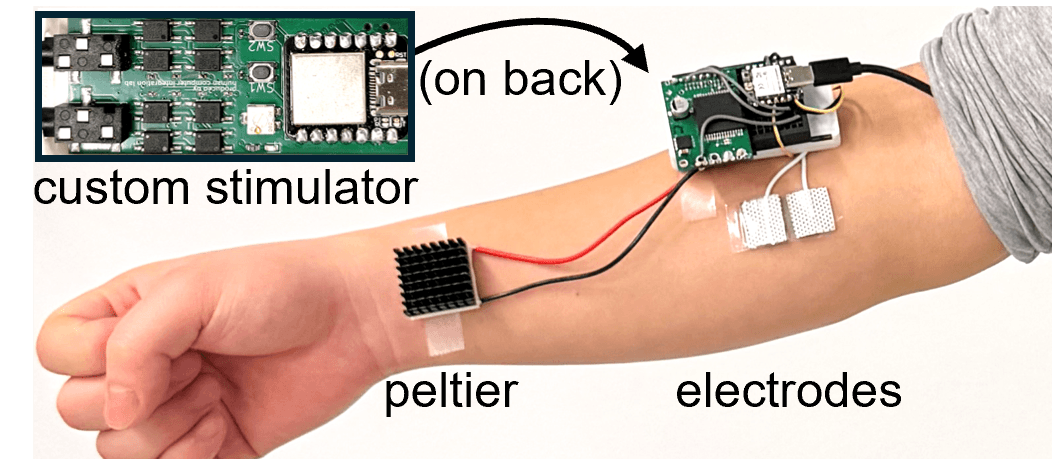}
  \Description{Photograph of the wearable hardware on the inner forearm. A Peltier module is near the wrist, gel electrodes are near the elbow, and the custom stimulator board is mounted on the back of the forearm.}
  \caption{Our hardware prototype attached to a user's arm.}
  \label{fig:hardware}
\end{figure}

Our main hardware prototype, depicted in Figure~\ref{fig:hardware}, was a wearable
version designed to trigger thermal referral via a combination of
peltier (thermal) and electrotactile stimulation (tactile). In the
study, participants wore only the electrodes and the peltier; the
remainder of the hardware shown in Figure~\ref{fig:hardware} sat on a table to minimize
attachment time and mitigate risk of wearables loosening up during
studies. To help readers replicate our design, we provide technical
details in an open-source repository\footnote{https://lab.plopes.org/\#electro-thermal-referral}.

\textbf{Logic control.} We developed a compact wearable prototype using
a WiFi-enabled \emph{ESP32-C3} microcontroller (\emph{Seeed}) to
coordinate a peltier element for thermal output and our custom-built
electrotactile stimulator for tactile output. The microcontroller
listens to WiFi OSC commands sent from VR applications (our
implementations run in \emph{Unity3D}) and activates the peltier and
electrotactile stimulator in real time, enabling thermal referral during
interactive VR experiences. Our device is powered by a power bank in the
user's pocket.

\begin{figure}[h!]
  \centering
  \includegraphics[width=\linewidth]{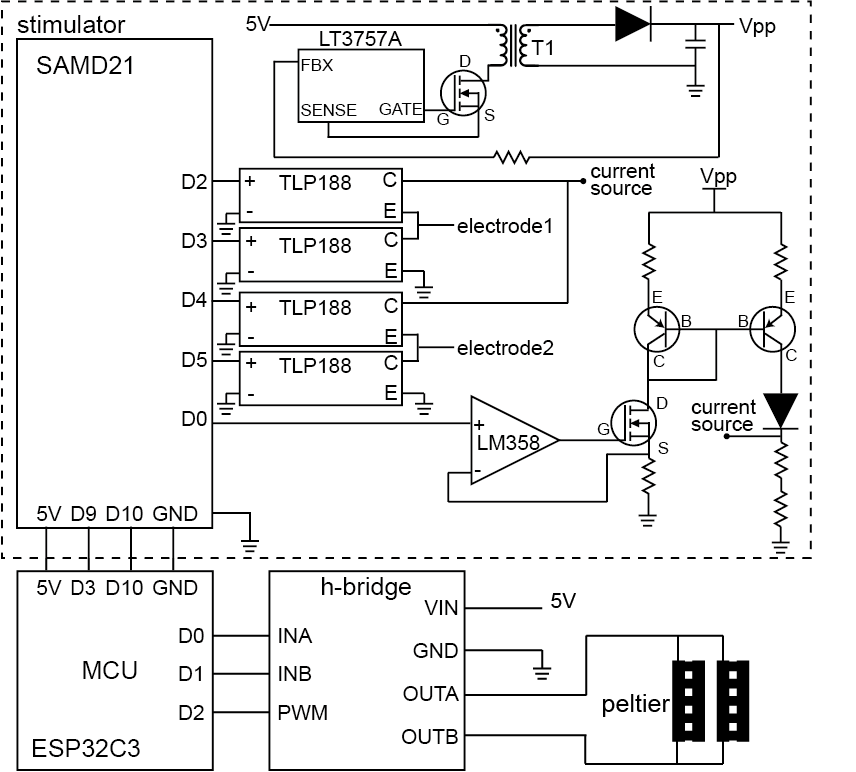}
  \Description{Circuit schematic of the wearable system. An ESP32-C3 drives a Peltier through an H-bridge, while an SAMD21, boost converter, current source, and optocoupler-based switches generate biphasic electrotactile stimulation.}
  \caption{High-level schematic of our device's circuitry.}
  \label{fig:circuit}
\end{figure}

\textbf{Thermal control.} The schematic representation of our
implementation is depicted in Figure~\ref{fig:circuit}. To drive our peltier element (25
$\times$ 25 mm) we utilize an h-bridge (\emph{ST VNH5019A}).

\textbf{Electrotactile control.} To generate electrotactile stimulation,
we implemented a custom and self-contained stimulator. At its core, our
stimulator is controlled by an \emph{SAMD21} microcontroller
(\emph{Seeed}) as it provides a DAC, which we utilize for programmatic
analog control of the stimulator's output current. The microcontroller
drives an \emph{LT3757A}-based boost converter that steps up the 5 V USB
supply to a high-voltage rail. The stimulation current is regulated by
an op-amp (\emph{LM358}) feedback loop that controls a high-voltage
MOSFET, establishing a precise reference current. This reference is
replicated to the output via a PNP current mirror (matched
\emph{FCX705TA} transistors), ensuring consistent delivery regardless of
impedance variations. The stimulation is routed to the electrodes
through optocoupler-based h-bridges (\emph{TLP188}), which implement
biphasic polarity switching to produce the AC square waveform required
for proper electrotactile stimulation \citep{Kaczmarek1991Electrotactile}. For additional safety,
the current is capped by a current limiting diode. Finally,
electrotactile stimulation is delivered through pre-gelled electrodes
placed on the forearm.

\hypertarget{overview-of-studies}{%
\section{Overview of Studies}\label{overview-of-studies}}

We conducted two user studies to validate the benefits of switching from
vibrotactile to electrotactile for thermal referral. To the best of our
knowledge, our studies are the first to explore the use of
electrotactile stimulation for thermal referral\emph{.} In our
\textbf{first study (psychophysics)}, we assessed how effective
electrotactile is in thermal referral compared to vibration. We found
that participants reported (1) a higher cold referral rate under
electrotactile; and (2) a higher thermal intensity in both hot and cold
conditions with electrotactile. In our second \textbf{study (VR
deployment),} participants experienced four different scenes
(non-contact heat, contact heat, non-contact cold and contact cold) and
compared vibration-based thermal referral to electrotactile-based
thermal referral. We found that using electrotactile stimulation,
participants rated significantly higher realism in non-contact heat,
non-contact cold and contact cold. Both of these studies were approved
by our local ethics committee (\emph{IRB22-0228}).

\hypertarget{experimental-apparatus}{%
\subsection{Experimental Apparatus}\label{experimental-apparatus}}

Both studies used our laboratory apparatus designed for precise stimulus
control. The apparatus shares all the core hardware from our wearable
prototype (see \emph{Implementation}) but, for stability and proper
control, was implemented in a stationary circuit (on the desk rather
than worn).

\textbf{Thermal stimulation:} As in our wearable prototype, thermal
stimulations are delivered via a peltier actuator (40 $\times$ 40 $\times$ 10 mm,
rated at 12 V). For additional stability and control of the cold
stimulation, we added a heat sink and cooling fan for heat dissipation,
as is typical of thermal research \citep{Mazursky2024ThermalGrasp}. The peltier was attached to
participants left forearm (ventral side) and secured with medical-grade
tape. The peltier was programmed to deliver cold (-5 $^\circ$C) or hot (+3 $^\circ$C)
stimuli relative to each participant\textquotesingle s baseline skin
temperature based in prior work \citep{Wang2024Thermal}; these temperatures were
measured by a thermistor (0.1 $^\circ$C precision) affixed to the peltier
surface, ensuring that heating/cooling power levels were held constant
across all trials. Heating rate was 5.5 $^\circ$C/s and cooling was 5.0 $^\circ$C/s
(measured at peltier via 0.1 $^\circ$C precision thermal camera, correlated via
thermistor).

\textbf{Tactile stimulation}: Unlike our wearable version, we added two
additional optocoupler and a bench-top power supply to drive similar
vibration motors as used in prior work \citep{Wang2024Thermal}, i.e., coin-type ERM
motors (10 mm diameter, 3 mm depth). Following the same parameters from
prior work \citep{Wang2024Thermal}, our ERM was driven at 2.68 V (hot) and 3.19 V
(cold), mounted directly on top of the gel electrodes to ensure
co-located stimulation across modalities. The electrotactile stimulation
was akin to that of our wearable device but used a desktop stimulator,
i.e., the more precise and medically-compliant \emph{Rehamove 3} (from
\emph{Hasomed}). Stimulations used a biphasic AC square waveform at 100
Hz with a 10 $\mu$s pulse-width. As is typical in electrotactile, frequency
was selected in pilot experiments, while pulse-width was set low to
allow per-user calibration by means of intensity adjustments (in
mA)---this intensity adjustment with short pulse-width was used to
ensure we can always induce weak electrotactile stimulations.
Stimulations were delivered via pre-gelled electrodes (40 $\times$ 20 mm).
Electrodes were attached to participants left forearm (ventral side) and
secured with medical-grade tape. Moreover, the distance between tactile
and thermal actuators varied based on the study: 8 cm in the first
psychophysics study (similar to that of prior work \citep{Wang2024Thermal}) and 12 cm
for our second study in VR (to best match the visuals in the virtual
experience).

\hypertarget{study-1-electrotactile-in-thermal-referral}{%
\section{\texorpdfstring{Study 1: Electrotactile in Thermal Referral
}{Study 1: Electrotactile in Thermal Referral }}\label{study-1-electrotactile-in-thermal-referral}}

Our first study builds on a psychophysics experiment from prior work
\citep{Wang2024Thermal} but introduces novel questions never previously investigated:
(1) \emph{can electrotactile stimulation induce thermal referral?}; and
(2) \emph{can lower intensity tactile stimulations} \emph{(e.g., just
above the feelable threshold) improve the quality of thermal referral?}

\textbf{Hypotheses}. Our hypotheses were that: (H1) electrotactile
stimulation improves thermal referral; and (H2) lower intensity tactile
stimulations can lead to purer thermal sensations, with less associated
tactile ``buzzing'' sensations.

\hypertarget{conditions}{%
\subsection{Conditions}\label{conditions}}

While our work follows a standard psychophysics experiment (modelled
after \citep{Wang2024Thermal}), it is the first to feature a systematic range of
conditions for thermal referral, i.e., rather than one, we have
\emph{five} distinct tactile conditions to isolate any possible
contributions of intensity and modality:

\textbf{Vibrotactile (abbreviated as \emph{vibro}):} A standard
suprathreshold (i.e., clearly feelable) vibration intensity based on
prior literature, i.e., we drove the ERM actuators at 2.68 V for hot
trials and 3.19 V for cold trials \citep{Wang2024Thermal}.

\textbf{Vibrotactile-minimum (\emph{vibro-min}).} A novel condition that
we explore for the first time in this work. Unlike the vibrotactile
baseline condition from prior work~\citep{Wang2024Thermal}, this \emph{vibro-min}
condition is the absolute detection threshold for vibration. This was
defined as the minimum voltage required to generate the first
perceptible sensation, constrained by the mechanical starting voltage of
motor (0.9~V). This value was determined per participant using a
standard 1-up 1-down psychophysics staircase procedure with five
reversals \citep{Cornsweet1962staircrase}, and the threshold was calculated as the mean of the
last three reversal points. The average threshold voltage across
participants in \emph{Study 1} was 0.96 V (SD=0.10).

\emph{\textbf{Electrotactile}}: Our novel electrotactile stimulation for
thermal referral was delivered at 100 Hz at 10 $\mu$s. To ensure this
condition was on equal footing with the \emph{vibro} baseline, we
perceptually matched its intensity with the \emph{vibro} condition via a
two-alternative forced choice intensity matching task: participants
received the standard vibro stimulus followed by an electrotactile
stimulus and judged which felt stronger. We used the 1-up and 1-down
staircase method to adjust the electrotactile intensity until the
participant perceived it at the same intensity as the vibrotactile. The
staircase terminated after five reversals, and the point of subjective
equality was the mean of the last three reversals. After calibration, we
found an average of 46.4 mA (SD=12.5) for the intensity-matched
electrotactile. Despite the findings that electrotactile alone can
produce weak cold sensations \citep{Saito2023Coldness,Saito2021Thermal}, none of our participants
reported feeling cold sensations during the extensive electrotactile
calibrations.

\textbf{Electrotactile-minimum (\emph{electrotactile-min})}: Another
novel condition, that applied our idea of tactile stimulation just above
the detection threshold but for electrotactile stimulation. This was
adjusted using the same psychophysics standard paradigm as the
\emph{vibro-min} condition and resulted in an average amplitude of 32.4
mA (SD=9.8).

\emph{\textbf{No-tactile}}: We added a final control condition where
participants feel only thermal stimulation. This was included to rule
out the possibility that passive tactile cues from the apparatus could
be sufficient to trigger thermal referral.

\hypertarget{trial-design}{%
\subsection{Trial design}\label{trial-design}}

Trials were largely modelled after \citep{Wang2024Thermal}, to ensure a direct
comparison with prior work and attempt a replication of their findings.
Thus, following prior work whenever possible, per-trial, participants
closed their eyes while thermal and tactile stimuli were delivered
simultaneously for 5 and 7 seconds, respectively (from prior work
\citep{Wang2024Thermal}). The tactile stimulus was intentionally extended by 2 seconds
to prevent perceptual confusion from the peltier\textquotesingle s slow
thermal recession. After the stimulation ended, participants answered
five questions.

The first three questions determined the referral state following the
classification established in \citep{Wang2024Thermal}: (Q1) did you feel a thermal
sensation (hot or cold)?; (Q2) identify the region of strongest thermal
sensation (choose between location of thermal or tactile actuator); and,
(Q3) was the thermal sensation also felt at the other location? As in
prior work, after a positive response to the first question, the answers
in the two latter questions classify each trial into one of four states:
\emph{No Referral} (Q2: thermal location, Q3: no), \emph{Weak Referral}
(Q2: thermal location, Q3: yes), \emph{Strong Referral} (Q2: tactile
location, Q3: yes), or \emph{Masking} (Q2: tactile location, Q3: no).

Additionally, to capture spatial and perceptual dimensions beyond binary
referral classification, we added two questions: (Q4) draw the perceived
thermal region; and (Q5) rate the perceived thermal vs. tactile
intensity of the trial (using continuous sliders to implement a visual
analog scale).

\hypertarget{study-procedure}{%
\subsection{Study procedure}\label{study-procedure}}

Our study procedure followed a within-subjects factorial design. This
resulted in 660 trials at 30 trials per participant: 2 temperatures (hot
\& cold) $\times$ 5 conditions $\times$ 3 repetitions. To prevent any confounding
factors, these 30 trials were then randomly shuffled, and participants
were not made aware of which condition they were exposed to in any
single trial. Finally, to ensure participants understood the
instructions, they completed 5 practice trials. Also, advancing to the
next trial was only allowed after our thermistor detected that
participant's skin temperature returned to baseline.

\hypertarget{participants}{%
\subsection{Participants}\label{participants}}

We recruited 22 participants (12 male, 10 female, average age = 26.8
years old, SD=2.9, all right-hand dominant). Participants received \$20
as compensation.

\hypertarget{results}{%
\subsection{Results}\label{results}}

\hypertarget{referral-rates-i.e.-how-often-thermal-referral-is-effective}{%
\subsubsection{\texorpdfstring{Referral rates (i.e., how often thermal
referral is effective)
}{Referral rates (i.e., how often thermal referral is effective) }}\label{referral-rates-i.e.-how-often-thermal-referral-is-effective}}

\begin{figure}[h!]
  \centering
  \includegraphics[width=\linewidth]{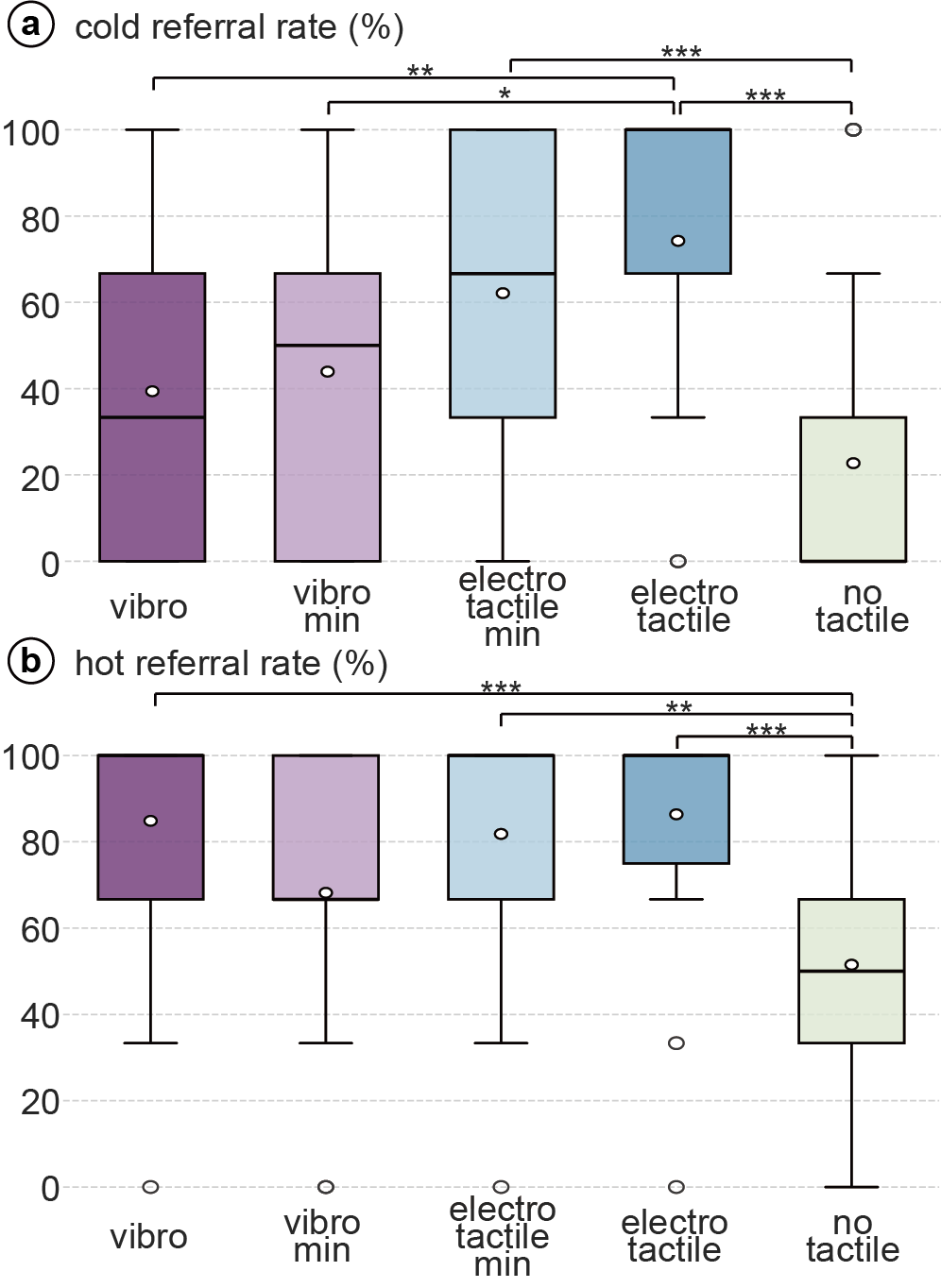}
  \Description{Box plots of cold and hot referral rates across vibro, vibro-min, electrotactile-min, electrotactile, and no-tactile conditions. Cold referral is higher for electrotactile than vibrotactile conditions, while hot referral remains high across tactile conditions.}
  \caption{Mean referral rates on five tactile conditions and two temperature conditions: cold (a) and hot (b). Asterisks depict *p\textless0.05, **p\textless0.01, ***p\textless0.001. Box plots show median, IQR (box), 1.5$\times$ IQR (whiskers), outliers, and mean.}
  \label{fig:referral-rate}
\end{figure}

Since referral rate takes discrete values, we adopted non-parametric
analysis using Friedman tests with Conover\textquotesingle s post-hoc
t-tests and Holm-Bonferroni corrections. This statistical analysis
revealed a significant main effect of tactile condition on referral
rates for both cold referral (\emph{$\chi^2$(4)}=26.88,
p\textless0.001) as depicted in Figure~\ref{fig:referral-rate} (a), and hot referral
(\emph{$\chi^2$(4)}=22.23, p\textless0.001) as depicted in
Figure~\ref{fig:referral-rate} (b).

\textbf{For cold stimuli,} post-hoc comparisons suggest that
transitioning from vibrotactile to electrotactile significantly
increased (p=0.005) referral rates from an average of 39\% (SD=37) with
\emph{vibro} condition to 74\% (SD=34) on the \emph{electrotactile}
condition. This result confirms our main hypothesis (H1) for cold
sensations, i.e., electrotactile improves cold thermal referral when
compared to vibrotactile. In fact, even against our novel
\emph{vibro-min} condition (i.e., just above perceptual threshold) there
was still a statistically significant improvement (p=0.018) from
switching from \emph{vibro-min} (M=44\%; SD=39) to
\emph{electrotactile}, further emphasizing the improved cold referral
rates induced by electrical stimulation. Critically, for cold events,
all forms of vibration (even our novel \emph{vibro-min}) failed to
significantly outperform our \emph{no-tactile} control (\emph{vibro} vs.
\emph{no-tactile}: p=0.212; \emph{vibro-min} vs. \emph{no-tactile}:
p=0.110), suggesting that adding vibration provided negligible benefits
over passive contact alone for cold referrals. Conversely, we did
observe a significant improvement (p\textless0.001) between
\emph{electrotactile-min} (M=62\%; SD=35) and \emph{no-tactile} (M=23\%;
SD=38).

\textbf{Conversely, for hot stimuli,} the referral rates shown in Figure
4 (b) remained stable at approximately 80\% regardless of the tactile
modality or intensity, with \emph{vibro-min} showing a marginal
decrease. In fact, statistical analysis found significant differences
between all experimental conditions (except \emph{vibro-min}) against
\emph{no-tactile} (M=52\%; SD=35), as follows: vibration (M=85\%; SD=27;
p\textless0.001), \emph{electrotactile-min} (M=82\%; SD=29; p=0.002),
\emph{electrotactile} (M=86\%; SD=27; p\textless0.001). However, in hot
referrals, no differences between the conditions themselves were found.
Interestingly, this result rejects the hypothesis (H1) in the case hot
referrals---unlike was the case in the cold referrals.

\hypertarget{results-per-referral-type-weak-vs.-strong-vs-masking}{%
\subsubsection{Results per referral-type (weak vs. strong vs,
masking)}\label{results-per-referral-type-weak-vs.-strong-vs-masking}}

\begin{figure}
  \centering
  \includegraphics[width=\linewidth]{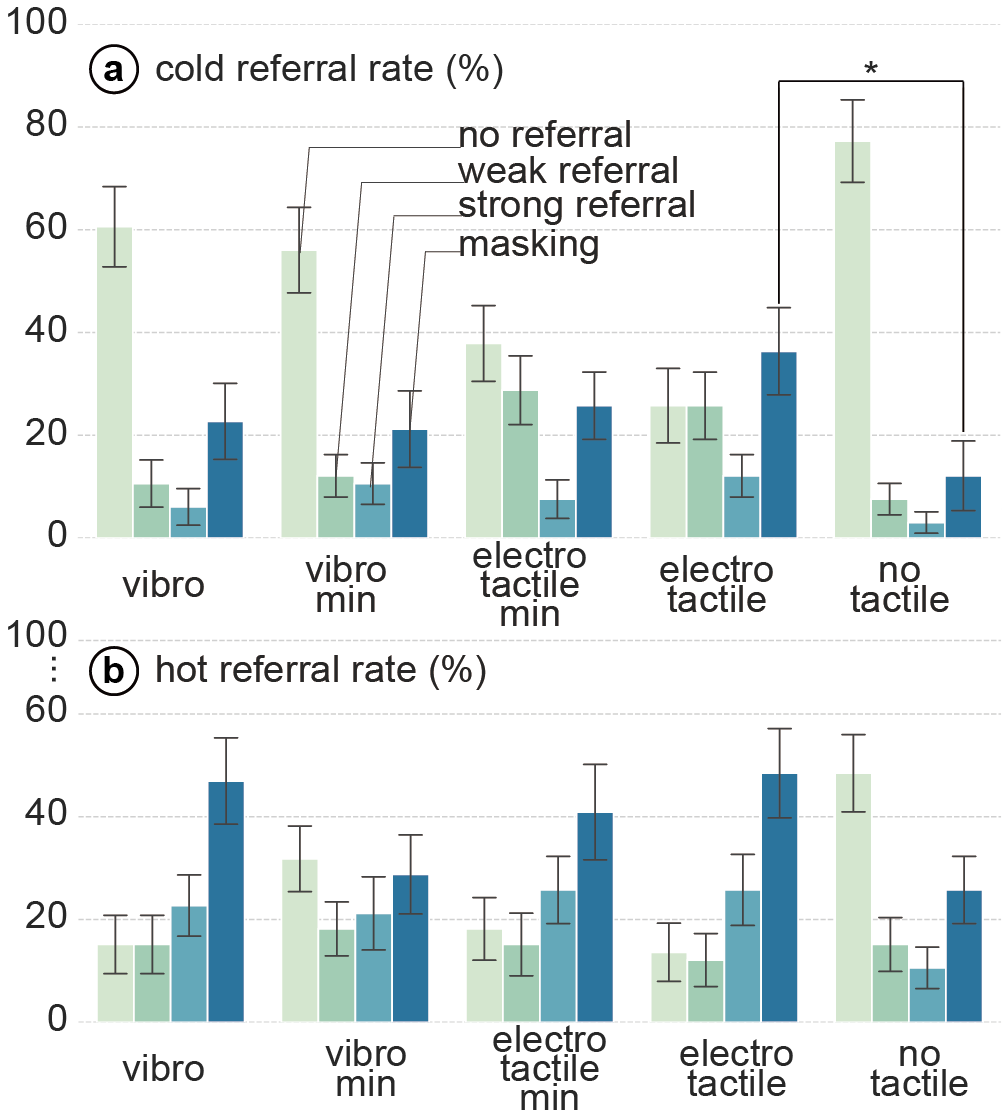}
  \Description{Grouped bar plots showing cold and hot referral outcomes classified as no referral, weak referral, strong referral, and masking across five tactile conditions.}
  \caption{Mean referral rates of no referral, weak referral, strong referral and masking on five different tactile conditions and two temperatures. Error bars depict standard errors. * depicts statistical significance at p\textless0.05.}
  \label{fig:referral-type}
\end{figure}

In Figure~\ref{fig:referral-type}, we break down results via the referral types following the
classification in \citep{Wang2024Thermal}: \emph{no-referral} (failed to elicit
thermals in the tactile location), \emph{weak referral} (referred
thermal, but weaker than at thermal location), \emph{strong referral}
(referred thermal, and stronger than at thermal location), and
\emph{masking} (all thermal at tactile location). This analysis follows
our previous statistical analysis (i.e., non-parametric analysis using
Friedman tests with Conover\textquotesingle s post-hoc t-tests and
Holm-Bonferroni corrections).

\textbf{For cold stimuli,} we observed a statistically significant
increase in masking rates ($\chi^2$(4)=11.32, p=0.023). With one significant
difference (p=0.011) between \emph{electrotactile} (M=36\%; SD=40) and
\emph{no-tactile} (M=12\%; SD=32). No other paired-comparisons exhibited
statistical differences. This further suggests, for the case of cold,
electrotactile results not only in a higher referral rate (our previous
analysis) but also in more masking (this analysis), adding more credence
to supporting H1, for cold referrals.

\textbf{Conversely, for hot stimuli,} and as in the previous analysis of
referral rates, no differences were found in referral types.

\hypertarget{corroborating-results-with-spatial-data-of-thermal-referrals}{%
\subsubsection{Corroborating results with spatial data of thermal
referrals}\label{corroborating-results-with-spatial-data-of-thermal-referrals}}

\begin{figure}
  \centering
  \includegraphics[width=\linewidth]{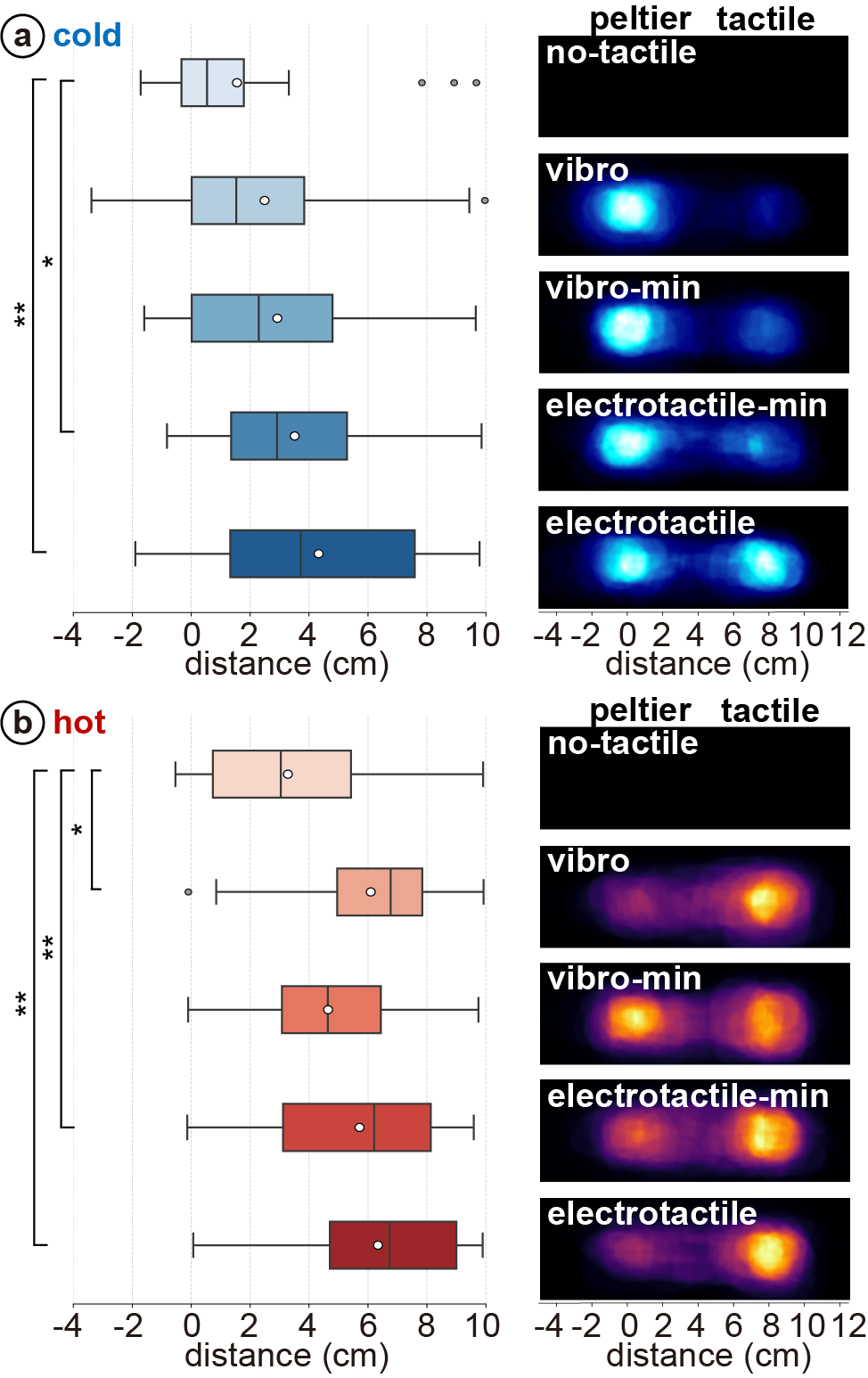}
  \Description{Spatial maps and box plots of perceived cold and hot locations. Color maps show participant drawings relative to the Peltier and tactile sites; box plots summarize horizontal centroid distances from the Peltier.}
  \caption{Distribution of perceived thermals by overlaying participant drawings. Red and blue hues represent hot and cold, and brightness encodes response frequency. Box plot shows the horizontal centroid distance relative to the peltier (0 cm). Asterisks depict *p\textless0.05, and **p\textless0.01. Box plots show median, IQR (box), 1.5$\times$ IQR (whiskers), outliers, and mean.}
  \label{fig:spatial}
\end{figure}

Insofar, our analysis was conducted following the procedure of prior
work (i.e., using the three questions that allow to detect a thermal
referral and classify its type). Next, we provide a spatial analysis of
\emph{where} participants felt the sensations, to further corroborate
our findings thus far. Figure~\ref{fig:spatial} depicts the spatial distribution of
perceived thermal sensations. We analyzed the results with a one-way
Repeated-Measures ANOVA with Greenhouse-Geisser corrections for
sphericity violations, followed by post-hoc t-tests with Holm-Bonferroni
corrections.

\textbf{For cold stimuli,} Figure~\ref{fig:spatial} (a) depicts a progressive shift in
participants' response density towards the tactile actuator as the
condition changes from \emph{no-tactile} to \emph{electrotactile}. To
quantify this effect, we analyzed the horizontal centroid distance
relative to the peltier (0 cm). Our ANOVA result revealed a significant
main effect of tactile condition on centroid location (F(3.00,
63.03)=5.60, p=0.002, \emph{$\eta^2$}\textsubscript{p}=0.21). Post-hoc
analysis confirmed a significant improvement (p=0.003) of
\emph{electrotactile} (M= 4.32 cm; SD= 3.59) over \emph{no-tactile}
(M=1.55 cm; SD=3.21). It also confirmed a significant
improvement (p=0.032) of \emph{electrotactile-min} (M=3.51 cm; SD=2.99)
over \emph{no-tactile}. Conversely, no significant differences were
found between all other conditions---including \emph{vibro} and/or
\emph{vibro-min} vs. \emph{no-tactile}---further reinforcing that
vibration was not effective at spatially shifting cold
sensations. This further corroborates our H1 for cold referrals.

\textbf{For hot stimuli,} Figure~\ref{fig:spatial} (b) depicts a lesser shift in
participants' response density towards the tactile actuator as the
condition changes from \emph{no-tactile} to \emph{electrotactile}. To
quantify this effect, we again analyzed the horizontal centroid distance
to the thermal actuator. The ANOVA result suggested a significant main
effect (F(3.00, 62.97)=8.05, p\textless0.001,
\emph{$\eta^2$}\textsubscript{p}=0.28). Post-hoc analysis confirmed a
significant improvement (p=0.015) of \emph{vibro} (M=6.09 cm; SD=2.91)
over \emph{no-tactile} (M=3.28 cm; SD=3.01). Moreover, it confirmed a
significant improvement (p=0.005) of \emph{electrotactile-min} (M=5.71
cm; SD=2.91) over \emph{no-tactile}. Also, it confirmed a third
significant improvement (p=0.005) of \emph{electrotactile} (M=6.34 cm;
SD=3.05) over \emph{no-tactile}.

\hypertarget{intensity-of-thermal-sensations}{%
\subsubsection{\texorpdfstring{Intensity of thermal sensations
}{Intensity of thermal sensations }}\label{intensity-of-thermal-sensations}}

Next, we analyze perceived thermal intensities, for both cold and hot
stimuli with the same ANOVA procedure (succinctly a repeated measures
with corrections for sphericity violations, and Holm-Bonferroni
corrected post-hoc t-tests).

\begin{figure}
  \centering
  \includegraphics[width=\linewidth]{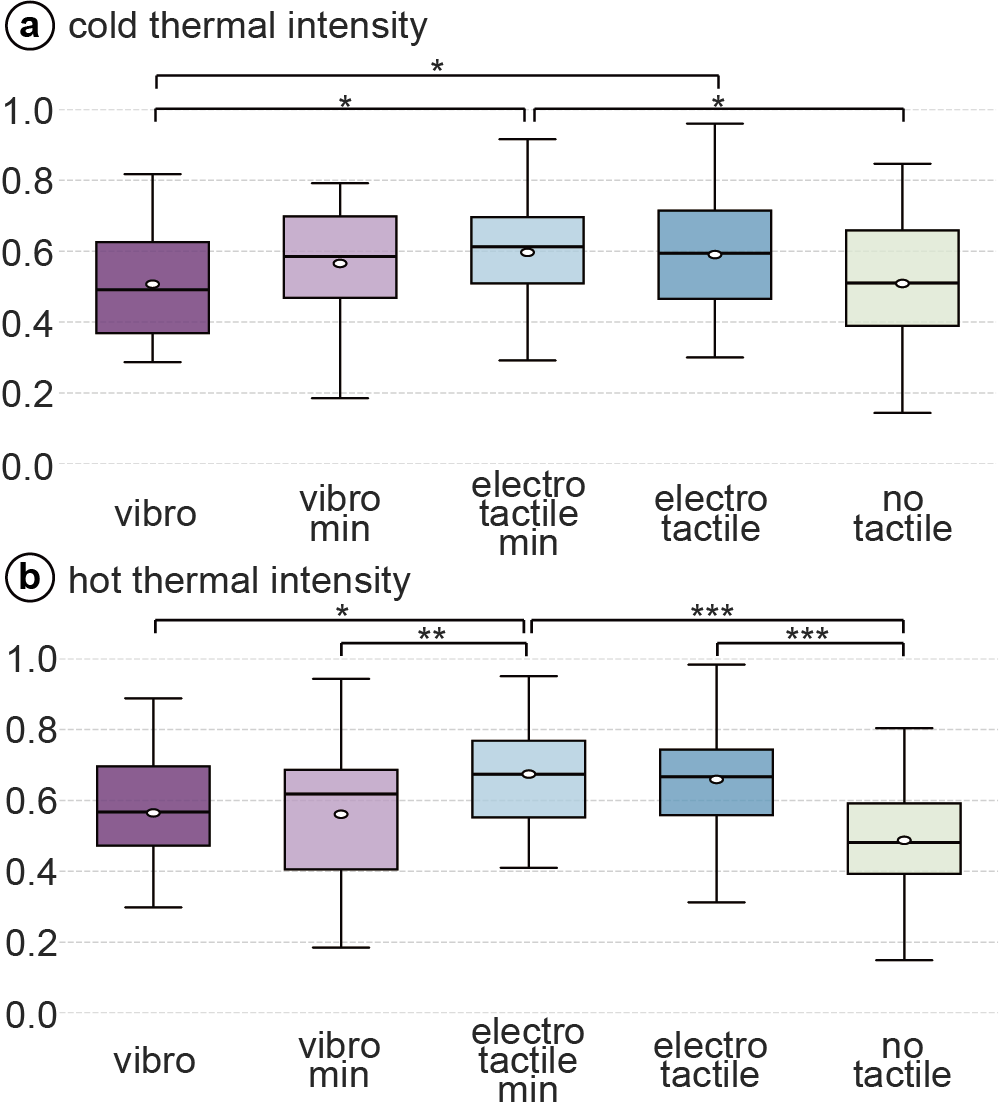}
  \Description{Box plots of perceived cold and hot thermal intensity for the five tactile conditions.}
  \caption{Perceived thermal intensity for (a) cold and (b) hot for different tactile conditions. Asterisks depict *p\textless0.05, **p\textless0.01, and ***p\textless0.001. Box plots show median, IQR (box), 1.5$\times$ IQR (whiskers), outliers, and mean.}
  \label{fig:thermal-intensity}
\end{figure}

\textbf{For cold stimuli,} Figure~\ref{fig:thermal-intensity} (a) depicts perceived cold intensity
for all five conditions. The ANOVA result suggested a significant main
effect of tactile condition (F(3.40, 71.32)=4.50, p=0.004,
$\eta^2$\textsubscript{p}=0.177). Post-hoc comparisons suggested a significant
improvement (p=0.023) of \emph{electrotactile-min} (M=0.60, SD=0.16)
over \emph{vibro} (M=0.51, SD=0.16). Importantly, it also confirmed a
significant improvement (p=0.018) of \emph{electrotactile} (M=0.59,
SD=0.18) over \emph{vibro}. Moreover, it also confirmed a third
significant improvement (p=0.035) of \emph{electrotactile-min} over
\emph{no-tactile} (M=0.51, SD=0.20).

\textbf{For hot stimuli}, Figure~\ref{fig:thermal-intensity} (b), ANOVA revealed a significant
main effect (F(2.95, 61.85) = 11.12, p \textless{} 0.001,
$\eta^2$\textsubscript{p} = 0.346). Post-hoc comparisons suggested a
significant improvement (p=0.015) of \emph{electrotactile-min} (M=0.67,
SD=0.15) over \emph{vibro} (M=0.56, SD=0.16). Moreover, it confirmed a
significant improvement (p=0.005) of \emph{electrotactile-min} over
\emph{vibro-min} (M=0.56, SD=0.20). It also confirmed a third
significant improvement (p\textless0.001) of \emph{electrotactile-min}
over \emph{no-tactile} (M=0.49, SD=0.16). Finally, it confirmed a fourth
significant improvement (p\textless0.001) of \emph{electrotactile}
(M=0.66, SD=0.15) over \emph{no-tactile}.

It is worth reminding that the peltier output power was held constant
across all conditions within each temperature. Despite this, the choice
of tactile modality significantly influenced perceived thermal
intensity. These results highlight a new advantage of electrotactile
(\emph{electrotactile-min} for hot/cold and \emph{electrotactile} for
hot) as it increases the perceived thermal intensity, compared to the
no-tactile (a statistical difference not observed in vibrotactile
conditions).

\hypertarget{purity-of-thermal-sensations}{%
\subsubsection{\texorpdfstring{Purity of thermal sensations
}{Purity of thermal sensations }}\label{purity-of-thermal-sensations}}

Next, we turn our analysis into the metric that directly addresses our
second hypothesis: \emph{do our new approaches} (firstly,
\emph{electrotactile}, but also the usage of \emph{weaker} tactile
stimulus) \emph{minimize the unwanted tactile percepts and maximize
thermal percepts?} In other words, \emph{do they improve the purity of
thermal sensations?}

To formally evaluate the dominance of the thermal sensation over the
tactile, we define a \emph{thermal purity} metric\emph{---}simply the
thermal intensity rating divided by the sum of the thermal and tactile
intensity ratings as follows:

\[\mathbf{Thermal\ purity =}\frac{\mathbf{Thermal\ intensity}}{\mathbf{Thermal\ intensity + Tactile\ intensity}}\]

To calculate this, we require thermal intensity (i.e., previous
analysis), and tactile intensity, which we analyze next.

\begin{figure}[h!]
  \centering
  \includegraphics[width=\linewidth]{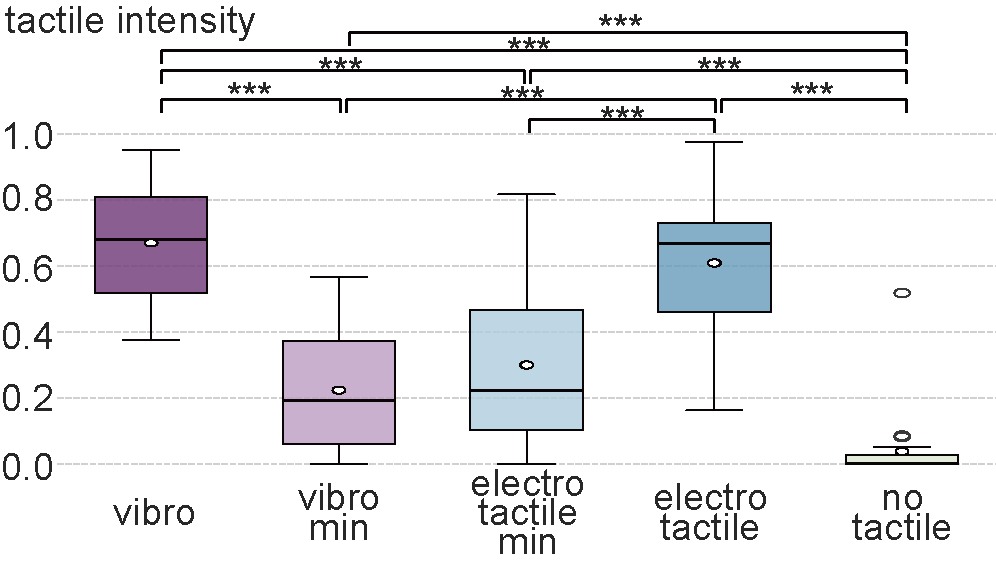}
  \Description{Box plot of tactile intensity for vibro, vibro-min, electrotactile-min, electrotactile, and no-tactile conditions, with significant pairwise comparisons indicated above the plot.}
  \caption{Tactile intensity for different tactile conditions. *** depicts statistical significance at p\textless0.001. Box plots show median, IQR (box), 1.5$\times$ IQR (whiskers), outliers, and mean.}
  \label{fig:tactile-intensity}
\end{figure}

\textbf{Tactile intensity.} We used a one-way Repeated-Measures ANOVA
with Greenhouse-Geisser corrections for sphericity violations, followed
by post-hoc t-tests with Holm-Bonferroni corrections, which suggested a
significant main effect (F(2.86, 60.01)=66.52, p\textless0.001,
\emph{$\eta^2$}\textsubscript{p}=0.76). Figure~\ref{fig:tactile-intensity} depicts these results across
all conditions. As expected, both \emph{vibro} vs. \emph{electrotactile}
as well as \emph{vibro-min} vs. \emph{electrotactile-min} were not found
to be significantly different with respect to their tactile intensity,
which validates the quality of our perceptual matching of the intensity
calibration. Conversely, all remaining pairwise comparisons were
statistically significant. First, as expected, \emph{vibro} (M=0.67; SD=
0.17) was found (p\textless0.001) to be stronger than stronger than
\emph{vibro-min} (M=0.22; SD=0.17), and \emph{electrotactile} (M= 0.61;
SD=0.21) was stronger (p\textless0.001) than \emph{electrotactile-min}
(M=0.30; SD=0.26). Again, this confirms that our apparatuses worked as
expected (i.e., the \emph{min} conditions were indeed correctly
calibrated). Moreover, we found \emph{vibro} to be stronger
(p\textless0.001) than \emph{electrotactile-min}. Also, we found
\emph{electrotactile} to be stronger (p\textless0.001) than
\emph{vibro-min}. Finally, when compared to \emph{no-tactile} (M=0.04;
SD=0.11), all conditions were stronger (p\textless0.001).

\begin{figure}
  \centering
  \includegraphics[width=\linewidth]{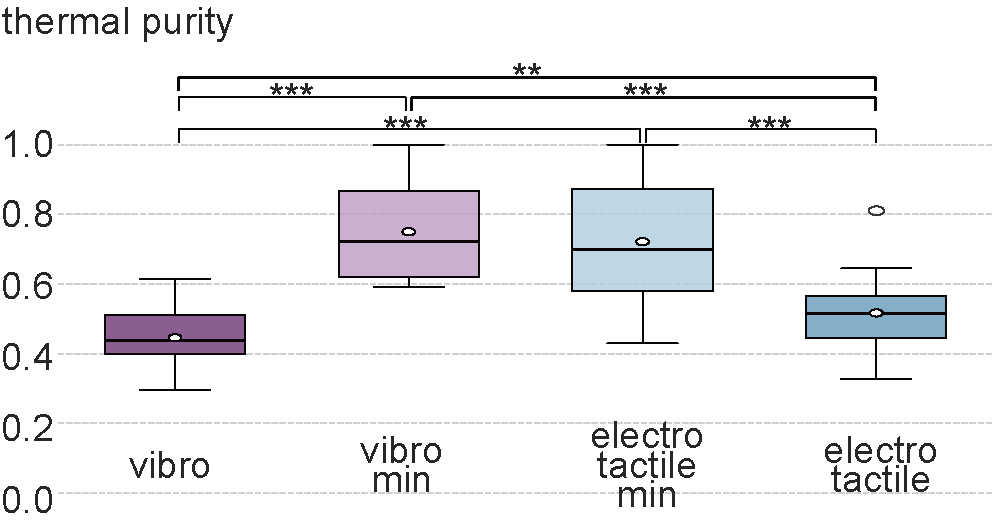}
  \Description{Box plot of thermal purity for vibro, vibro-min, electrotactile-min, and electrotactile conditions, with significant pairwise comparisons.}
  \caption{Thermal purity across all experimental conditions (except no-tactile, which does not include any tactile intensity, thus saturates thermal purity). Asterisks depict **p\textless0.01, ***p\textless0.001. Box plots show median, IQR (box), 1.5$\times$ IQR (whiskers), outliers, and mean.}
  \label{fig:thermal-purity}
\end{figure}

\textbf{Thermal purity.} As this data satisfied normality, we used a
one-way Repeated-Measures ANOVA with Greenhouse-Geisser corrections for
sphericity violations, followed by post-\emph{hoc} t-tests with
Holm-Bonferroni corrections, which revealed of a main effect of
condition (F(2.20, 46.11) = 36.36, p\textless0.001,
$\eta^2$\textsubscript{p}=0.634). Figure~\ref{fig:thermal-purity} depicts the calculated thermal
purity across all conditions (except the \emph{no-tactile}, which was
not included in this analysis, since it is a control, not a referral
condition). As expected, given the low tactile ratings of
\emph{vibro-min} and \emph{electrotactile-min}, we found no significant
difference in their thermal purity.

Four pair-wise comparisons are directly responsible in answering our
secondary hypothesis (H2---weaker tactile conditions improve thermal
purity). First, we found that \emph{vibro-min} (M=0.75; SD=0.14)
displayed a significant higher thermal purity (p\textless0.001) when
compared to both \emph{vibro} (M=0.45; SD=0.08) and
\emph{electrotactile} (M=0.52; SD=0.11). Second, we found that
\emph{electrotactile-min} (M=0.72; SD=0.18) displayed a significant
higher thermal purity (p\textless0.001) when compared to both
\emph{vibro} and \emph{electrotactile}. Taken together, these
statistical results confirm our second hypothesis, as both conditions
with minimum tactile intensity (just above feelable threshold) maximized
perceive thermals and minimized perceived tactile (i.e., more thermally
``pure''). Finally, we also found that \emph{electrotactile} was
perceived at a higher thermal purity (p=0.003) than \emph{vibro}.

\hypertarget{discussion-synthesizing-insights-from-study-1}{%
\section{Discussion: Synthesizing insights from data}\label{discussion-synthesizing-insights-from-study-1}}

\textbf{Summary of findings.} We validated two novel proposals: (1)
electrotactile stimulation can outperform vibrotactile for cold
referrals; and (2) weaker tactile stimulation (both for electrotactile
and vibrotactile) can improve the purity of thermal sensations,
mitigating some of the unwanted tactile sensation. Moreover, we also
provided a replication of the insights of prior work on thermal referral
\citep{Wang2024Thermal}, further emphasizing the promise that this technique holds.

\textbf{Discussion of possible mechanism.} Our main result, which is the improvement
of cold referral rates, is impactful as this is a well-documented
challenge, i.e., the asymmetric rates of cold being approximately 20\%
lower than hot \citep{Wang2024Thermal}. While the prevailing hypothesis attributes this
to thermoreceptor distribution \citep{Ho2011Mechanisms,Luo2020High}, our findings suggest this is
incomplete: if the asymmetry were driven solely by thermoreceptor
distribution, changing the tactile modality should have little to no
effect. Yet switching to electrotactile boosted cold referral from 39\%
to 74\%, while hot referral remained stable at around 85\%. Likely,
there is another underlying mechanism for cold sensations that remains
elusive, which is activated by electrotactile. We discuss the following
possible explanations: (1) tactile attention: electrotactile activates a
broader range of mechanoreceptors than vibration \citep{KajimotoElectro}; (2) beyond
mechanoreceptors: electrotactile stimulation may activate myelinated A-$\delta$
afferents that contribute to cold sensation \citep{Rubinstein1991Analytical,Saito2023Coldness,Saito2021Thermal}. (3) Hybrid
explanations: multiple mechanisms can contribute partially; and (4)
undiscovered mechanisms. 

\textbf{No-tactile caution.} We also found new and unexpected insights
since our experiment was one of the few to feature a \emph{no-tactile}
condition (\citep{Cataldo2016Thermal} also uses no-tactile, but it uses no active
actuator---only passive haptics via a piece of glass). In it, we noted
that during thermal referrals, illusory referrals (no-tactile condition)
can be surprisingly common (at 23\% for cold referral rate and 52\% for
hot). This finding was not previously reported in HCI literature but is
important for anyone utilizing this illusion. There are two possible
factors: the mechanical pressure exerted by the electrodes or vibration
motors may provide sufficient somatic input to trigger a weak illusion
or---albeit less likely---expectation effects inherent from the
randomized experimental design.

\begin{figure}[h!]
  \centering
  \includegraphics[width=\linewidth]{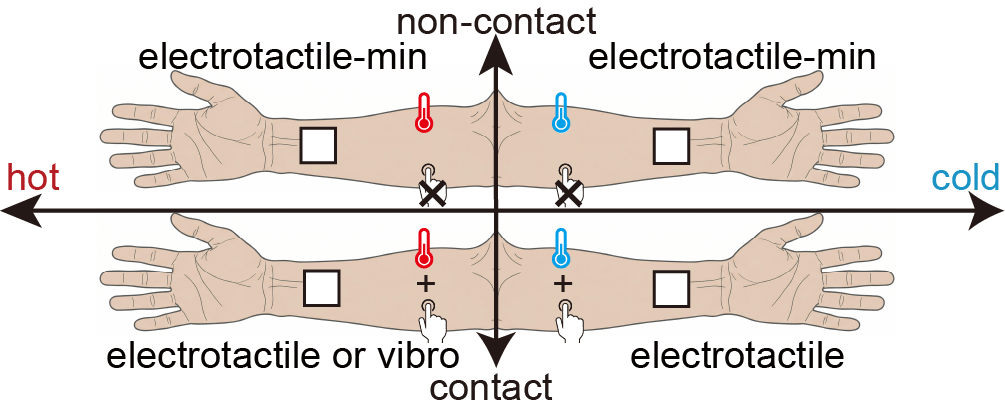}
  \Description{Conceptual two-axis guide for selecting tactile modalities by intended thermal temperature and contact level. Electrotactile-min is positioned for non-contact hot and cold events, while stronger electrotactile is positioned for contact events.}
  \caption{Conceptual guide to modality selection (\emph{vibrotactile}, \emph{electrotactile}, or \emph{min} conditions for weak stimulation).}
  \label{fig:guidelines}
\end{figure}

\textbf{Guidelines.} Next, a surprising part of our findings is that our
results do not suggest merely swapping electrotactile for vibrotactile,
but rather that these modalities can be used depending on target
sensations. Figure~\ref{fig:guidelines} synthesizes these results into guidelines by
depicting how modality can be chosen based on two perceptual dimensions:
intended tactile sensation (weak vs. strong) and intended thermal
sensation (hot vs. cold). As such, one hypothesizes that distinct
thermal experiences might benefit from different tactile actuators. For
non-contact heat events, \emph{electrotactile-min} emerges as a possible
choice as it maximizes thermal purity (minimizing tactile sensations)
and sustains a high referral rate. For contact heat events, both
standard \emph{electrotactile} and \emph{vibro} are highly effective,
yielding robust referral rates and providing the tactile feedback
necessary to simulate physical contact with a hot object. For
non-contact cold events, \emph{electrotactile-min} might be the most
suited modality as it provides a higher referral rate to trigger the
challenging cold referral illusion while maintaining the high thermal
purity expected of a non-contact event. Finally, for contact cold
events, \emph{electrotactile} emerges as a possible choice as it
achieves the highest overall cold referral rate and spatial
displacement, while simultaneously delivering the required tactile
feedback to simulate touching a cold surface. If validated, which is the
aim of our second study, such guidelines could enable future researchers
\& designers to optimize thermal referrals.

\hypertarget{study-2-rate-the-sensation-reality-in-vr}{%
\section{Study 2: Rate the sensation reality in
VR}\label{study-2-rate-the-sensation-reality-in-vr}}

Despite the electrotactile benefits and modality-selection guidelines we
found in Study 1, it could not prove this leads to measurable benefits
outside of a controlled psychophysics experiment, so we designed Study 2
to evaluate whether these advantages translate to improved realism in
VR.

\textbf{Hypothesis.} Our main hypothesis (H3) was that our modality
selection guidelines (from Study 1) would yield significantly higher
perceived realism across VR thermal scenarios, than the vibrotactile
approach (baseline).

\hypertarget{participants-1}{%
\subsection{Participants}\label{participants-1}}

We recruited 20 participants (11 male, 9 female, average age=25.9 years
old, SD=3.7, two left-handed). Participants gave informed consent and
received \$20 as compensation.

\hypertarget{thermal-tasks-conditions.}{%
\subsection{Thermal tasks \&
conditions.}\label{thermal-tasks-conditions.}}

\begin{figure}[h!]
  \centering
  \includegraphics[width=\linewidth]{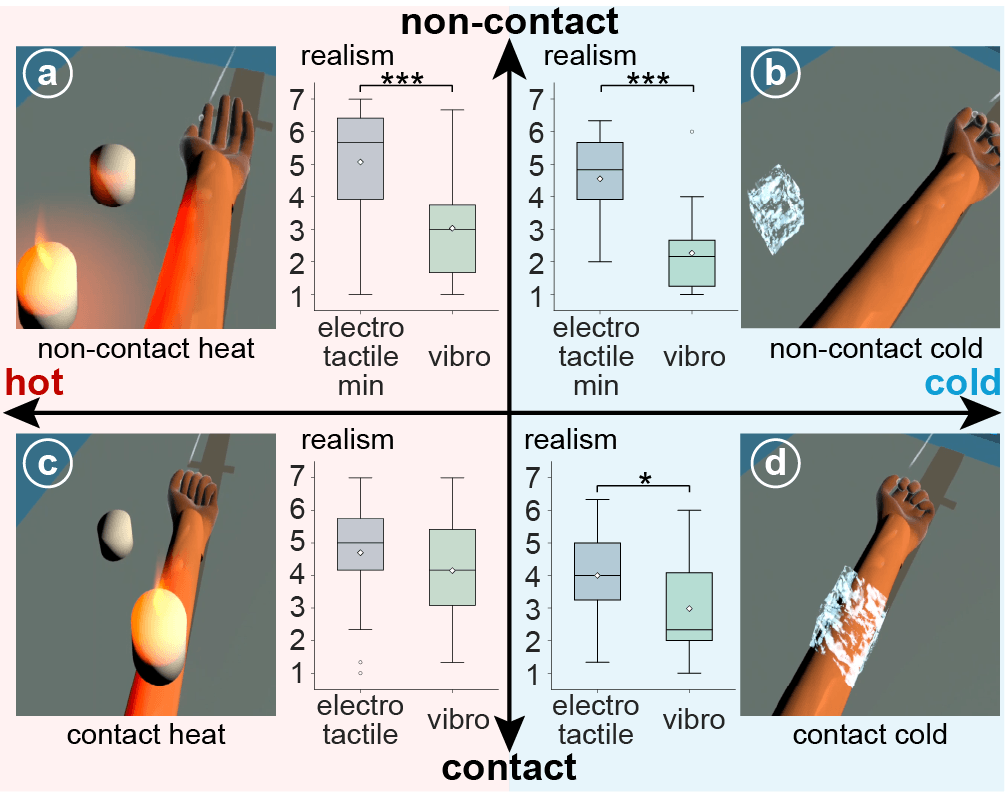}
  \Description{Four VR scenarios comparing realism ratings for electrotactile and vibrotactile thermal referral: non-contact heat, non-contact cold, contact heat, and contact cold.}
  \caption{Realism in four thermal VR tasks. Asterisks depict *p\textless0.05, and ***p\textless0.001. Box plots show median, IQR (box), 1.5$\times$ IQR (whiskers), outliers, and mean.}
  \label{fig:vr-realism}
\end{figure}

We designed four interactive tasks in VR (Figure~\ref{fig:vr-realism}) that directly map
to the four distinct thermal-tactile experiences using the two
dimensions shown in Figure~\ref{fig:guidelines}: (X) temperature and (Y) contact. In these
VR experiences, all animation timings were kept consistent; only visuals
for hot and cold were changed.

\textbf{Non-contact heat}: participants observed two virtual candles
while resting their avatar's arm on a virtual table: one near their
wrist (real heat) and one near their elbow (referred heat). The candles
were activated one at a time in a timed sequence. Participants
experienced this task under two conditions: \emph{vibro} (baseline) and
\emph{electrotactile-min}.

\textbf{Contact heat:} participants observed two candles and judged
their realism: one near their wrist (real heat) and one sitting on top
of their arm, near their elbow (referred heat), simulating physical
contact with their skin. Participants experienced this under
\emph{vibro} vs. \emph{electrotactile}.

\textbf{Non-contact cold:} participants observed two virtual ice cubes
and judged their realism: one near their wrist (real cold) and one near
their elbow (referred cold). Participants experienced this under
\emph{vibro} vs. \emph{electrotactile-min}.

\textbf{Contact cold:} participants observed the same ice cubes and
judged their realism: one near their wrist (real cold) and one near
their elbow and sitting on top of their arm (referred cold), simulating
physical contact with their skin. Participants experienced this under
\emph{vibro} vs. \emph{electrotactile}.

\hypertarget{study-procedure-1}{%
\subsection{Study procedure}\label{study-procedure-1}}

We performed our previous calibration to equalize the perceived
intensity of vibration and electrotactile conditions, resulting in an
average \emph{electrotactile-min} intensity of 30.1 mA (SD=8.2) and
\emph{electrotactile} of 49.3 mA (SD=11.8). Next, we established
perceptual anchors for the realism scale, by letting participants feel
examples of real thermal+tactile sensations using a peltier attached to
either a prop candle (safe to touch, no flame burn) and a prop ice cube
(safe to touch, no cold burn). We employed these anchors since it has
been established that VR visuals can bias the perceived thermal location
\citep{Gunther2024Assessing}.

Participants completed the four tasks in randomized order. Each
comprised three repetitions per condition (\emph{vibrotactile} vs.
\emph{electrotactile} or \emph{electrotactile-min}). Per trial,
participants rated the realism of the thermal experience associated with
the second virtual object on a 7-point Likert scale (1 = not realistic,
7 = perfectly realistic). To mitigate sequence effects, we shuffled all
six trials inside each task and participants were not told which
condition a trial belonged to. This resulted in 480 trials = 4 tasks $\times$ 2
conditions $\times$ 3 repetitions $\times$ 20 participants. At the end, participants
were offered a chance to provide verbal feedback.

\hypertarget{results-1}{%
\subsection{Results}\label{results-1}}

To statistically analyze the ordinal Likert data for perceived realism,
we employed Wilcoxon Signed-Rank tests with Holm-Bonferroni corrections
for multiple comparisons. Figure~\ref{fig:vr-realism} depicts realism ratings in all four
tasks.

\textbf{For non-contact heat events,} depicted in Figure~\ref{fig:vr-realism} (a),
participants rated \emph{electrotactile-min} (M=5.1; SD=1.8) to be more
realistic (p\textless0.001) than \emph{vibro} (M=3.0; SD=1.5).

\textbf{For non-contact cold events,} depicted in Figure~\ref{fig:vr-realism} (b),
participants rated \emph{electrotactile-min} (M=4.6, SD=1.3) to be more
realistic (p\textless0.001) than \emph{vibro} (M=2.3, SD=1.2).

\textbf{For contact hot events,} depicted in Figure~\ref{fig:vr-realism} (c)---as
expected---we did not see a significant different between participants
ratings on \emph{electrotactile} (M=4.7; SD=1.7) and \emph{vibro}
(M=4.2; SD=1.8)---consistent with Study 1's findings.

\textbf{For contact cold events,} depicted in Figure~\ref{fig:vr-realism} (d),
participants rated \emph{electrotactile} (M=4.0, SD=1.3) to be more
realistic (p=0.031) than \emph{vibro} (M=3.0, SD=1.6).

Post-study commentaries corroborated these results. Nine out of 20
participants explicitly identified the ``vibrotactile buzz'' as a source
of perceptual conflict, noting that the vibration caused their arm to
shake in a manner inconsistent with the expected interaction. Two
participants noted that the spatial correspondence between the virtual
object and the perceived thermal sensation was more accurate under
electrotactile, suggesting that electrotactile's increased realism stems
not only from reduced tactile distraction but also from improved
visuo-haptic spatial congruence.

In summary, we confirmed our hypothesis (H3) that electrotactile
significantly improves realism over \emph{vibro} in three of four VR
scenarios, suggesting promise in our modality selection guidelines.

\hypertarget{demonstrating-our-insights-in-a-vr-example}{%
\section{Demonstrating our insights in a VR
example}\label{demonstrating-our-insights-in-a-vr-example}}

To demonstrate the practical applicability of our modality selection
guidelines, we developed an ``iced tea making'' VR experience (Figure
12) that switches between configurations to best render both contact and
non-contact thermal events.

\begin{figure}[h!]
  \centering
  \includegraphics[width=\linewidth]{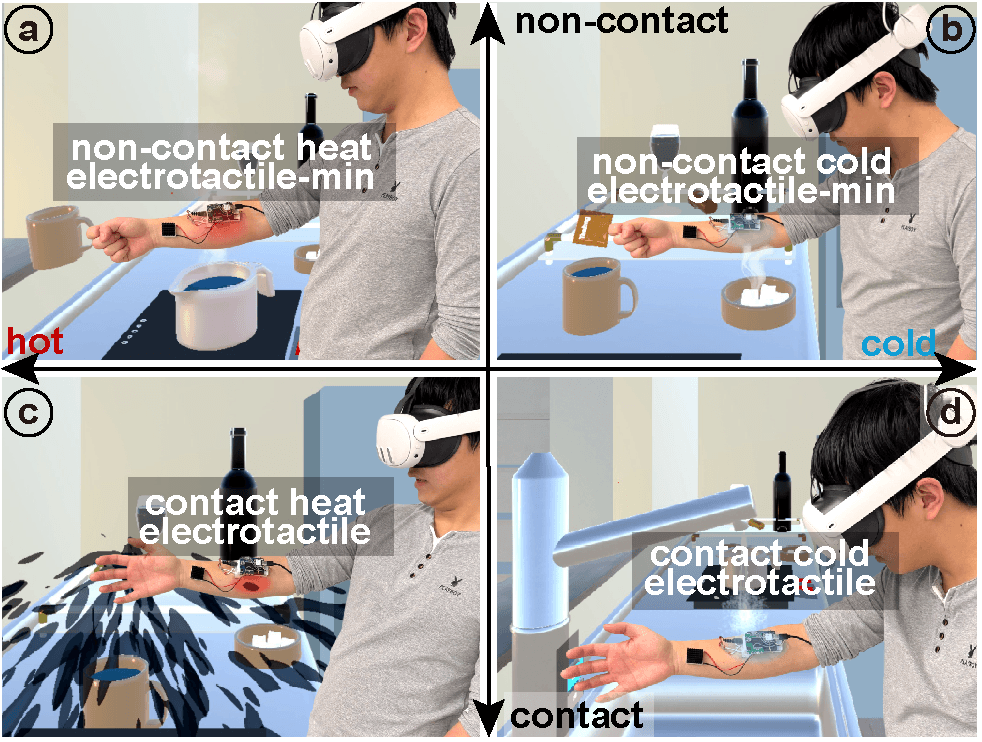}
  \Description{Four-panel mixed-reality image sequence of an iced-tea-making task. The panels show non-contact heat, non-contact cold, contact heat, and contact cold events rendered with electrotactile thermal referral.}
  \caption{Using electrotactile-based thermal referral to feel both tactile and thermal sensations in VR environment.}
  \label{fig:vr-demo}
\end{figure}

First, in Figure~\ref{fig:vr-demo} (a), as the user reaches past a boiling pot to fetch
a cup, steam diffuses onto their arm---a non-contact heat event is
rendered via \emph{electrotactile-min} to deliver warmth without much
tactile contact. Then, in Figure~\ref{fig:vr-demo} (b), while placing a tea bag, the
arm hovers over ice cubes and \emph{electrotactile-min} drives cold
referral to recreate the ambient chill without much physical touch.
Suddenly, in Figure~\ref{fig:vr-demo} (c) hot water splashes onto the arm---a contact
heat event rendered via \emph{electrotactile}, providing strong thermal
and tactile feedback. Finally, in Figure~\ref{fig:vr-demo} (d), the user washes their
scalded arm under a cold faucet as \emph{electrotactile} delivers cold
and tactile sensations for this contact cold event.

\section{Limitations}

Finally, as the first exploration of this novel method to achieve
thermal referral, our investigation is not without limitations, which
reveal future research directions: (1) with electrotactile stimulation,
thermal referral of heat sensations remains similar to that of
vibrotactile; (2) neither our approach nor vibrotactile-based referral
achieve a 100\% referral rate---calling for more research in this area,
both in HCI but also in neuroscience; (3) similarly, as with
vibration-based referrals, the referral type (e.g., weak, strong, or
masking) is also variable; (4) like the vast majority of prior work in
this area, we only focused on the arm; (5) we employed a fixed distance
between tactile and thermal actuators and did not investigate the
distance effect on referral; and finally, (6) as in prior work \citep{Wang2024Thermal},
our study also focused on static thermals; extending to moving thermal
sensations remains an open direction, though we see no fundamental
barrier in principle.

\hypertarget{conclusions-and-future-work}{%
\section{Conclusions and future
work}\label{conclusions-and-future-work}}

In this paper, we proposed electrotactile as a method for realizing
thermal referral. Our results demonstrate that electrotactile
stimulation significantly improves cold referral rates, mitigates
existing limitations with non-contact thermal events, and enhances
perceived realism across diverse VR scenarios. Our empirical results
were synthesized into guidelines, which we validated in our second
study, that might enable future researchers \& designers to optimize
thermal referrals in interactive domains, such as AR/VR.

Future work
might opt to explore additional body sites and variable actuator
distances for generalizability, investigating long-term comfort during
extended use, and studying how electrotactile parameters (e.g.,
frequency, pulse width) can be further optimized to improve referral
quality. As researchers has shown that different thermal rates influence
thermal sensations \citep{Filingeri2016Neurophysiology,Kenshalo1968Warm}, we will also study how different rates
influence thermal referral. We believe that the shift from vibrotactile
to electrotactile stimulation opens new design possibilities for richer
and more expressive thermal experiences in virtual and augmented
reality.

\begin{acks}
We thank Prof.\ Jasmine Lu for proofreading assistance; Yun Ho for
assistance with the figures and video; Prof.\ Jas Brooks for discussions
during the early stages of this investigation; and, especially, Romain Nith
for insightful discussions and for composing the original song for the video
figure.

This material is based upon work supported by the National Science Foundation
under Grant No.\ 2047189 and 2212352. Any opinions, findings, conclusions,
or recommendations expressed in this material are those of the authors and do
not necessarily reflect the views of the National Science Foundation.
\end{acks}

\bibliographystyle{ACM-Reference-Format}
\bibliography{references}

\end{document}